\documentclass[twocolumn,showpacs,preprintnumbers,amsmath,amssymb]{revtex4}

\usepackage{graphicx}
\usepackage{dcolumn}
\usepackage{bm}

\begin{document}


\title{Tensor  $A_{yy}$ and vector $A_{y}$  analyzing powers 
in the $H(d,d^\prime)X$ and  
$^{12}C(d,d^\prime)X$  reactons
at initial deuteron momenta of 9 GeV/$c$ in the region of baryonic resonances 
excitation.}

\author{
V.P.~Ladygin{\footnote{Electronic address:~ladygin@sunhe.jinr.ru}}, 
L.S.~Azhgirey,
S.V.~Afanasiev, V.V.~Arkhipov,
V.K.~Bondarev, Yu.T.~Borzounov, 
L.B.~Golovanov,
A.Yu.~Isupov, V.I.~Ivanov, A.A.~Kartamyshev,
V.A.~Kashirin, A.N.~Khrenov, V.I.~Kolesnikov,
V.A.~Kuznetsov,  N.B.~Ladygina, A.G.~Litvinenko,
S.G.~Reznikov,  P.A.~Rukoyatkin, A.Yu.~Semenov, I.A.~Semenova,
G.D.~Stoletov, A.P.~Tzvinev,  
V.N.~Zhmyrov,  L.S.~Zolin}

\affiliation{%
Joint Institute for Nuclear Research, Dubna, Russia \\
}%

\author{G.~Filipov}
\altaffiliation[Also at ]{Joint Institute for Nuclear Research, Dubna, Russia}
\affiliation{
Institute of Nuclear Research and Nuclear Energy, Sofia, Bulgaria\\
}%

\date{\today}

\begin{abstract}
The angular dependence of the tensor $A_{yy}$ and vector $A_{y}$ analyzing powers in the
inelastic scattering of deuterons with a momentum  of 9.0 GeV/$c$ on
hydrogen and carbon 
have been measured.
The range of measurements corresponds to the 
baryonic resonance excitation with  masses $\sim$2.2--2.6~GeV/$c^2$. 
The  $A_{yy}$ data being
in  good agreement with the
previous results demonstrate an
approximate  $t$ scaling up to -1.5 (GeV/$c$)$^2$.
The large values of $A_y$ show a significant
role of the spin-dependent part of the elementary amplitude of the 
$NN\to NN^*$ reaction.
The results of the experiment are compared with model
predictions of the plane-wave impulse approximation.
\end{abstract}

\maketitle

\section{\label{sec:intro} Introduction}

Inelastic scattering of polarized deuterons on hydrogen and nuclei
at high energies has been investigated during the last decade in Dubna 
\cite{45_55}--\cite{lad3} and Saclay \cite{morlet,ljuda}.
The main goal of these studies is to investigate the properties 
of the baryonic resonances via measurements of polarization
observables in the $(d,d^\prime)X$ reaction.

Since the deuteron is an isoscalar probe,
inelastic scattering of deuterons on hydrogen,
${\rm H}(d,d')X$, is selective  to the isospin $1/2$ and
can be used to obtain  information on the
formation of baryonic resonances $N^*(1440)$,
$N^*(1520)$, $N^*(1680)$, and others.
The measurements of analyzing powers, polarizations of the
scattered deuterons and different polarization transfers could   
allow one to obtain the spin-flip probabilites of the $(d,d^\prime)$ reaction sensitive to the
quantum numbers of baryonic resonances. 
The set of observables for the reconstruction of 
the spin-flip probabilities in the $A(d,d^\prime)A^*$ has been proposed in ref.\cite{suzuki}.
Such an experiment has been realized at RIKEN \cite{satou} at 270 MeV to
study the excitation levels of ${\rm ^{12}C}$.

At the same time, inelastic scattering of deuterons
on nuclei at high transferred momenta
can be considered as a
complementary method to elastic
$pd$- and $ed$-scattering, deuteron breakup reaction,
electro- and photodisintegration of the deuteron
to investigate the deuteron spin structure at
short distances and to search for the manifestation of
non-nucleonic degrees of freedom. 

However, due to lack of the experimental techniques to measure the polarization of 
the high-energy deuterons, only tensor and vector analyzing powers 
in the $(d,d^\prime)X$ reaction have been obtained in the vicinity of the baryonic
resonances excitation.

The tensor analyzing
power $T_{20}$ in the vicinity of the Roper resonance ($P_{11}(1440)$)
excitation  has been measured 
on hydrogen and carbon targets at Dubna
\cite{45_55} and on
hydrogen  target at Saclay \cite{morlet}.
The measurements of $T_{20}$
in the deuteron scattering at 9 GeV/$c$ on hydrogen and carbon
have been performed for missing masses up to $M_X\sim 2.2$
GeV/$c^2$ \cite{azh9}. The experiments have shown a large negative
value of $T_{20}$ at momentum transfers of $t\sim -0.3$ (GeV/$c$)$^2$.
Such a behaviour of the tensor analyzing power has been interpreted
in the framework of the $\omega$-meson exchange model \cite{egle1} as due
to the longitudinal isoscalar form factor of the Roper resonance
excitation and deuteron form factors $t$-dependences \cite{egle2}. 
The measurements of the tensor and vector
analyzing powers $A_{yy}$ and $A_y$ at  9 GeV/$c$ and 85 mrad of the
secondary deuterons emission angle in the vicinity of the
undetected system mass of $M_X\sim 2.2$ GeV/$c^2$ \cite{lad1} have shown
large values. The obtained results are in satisfactorily agreement with the
plane wave impulse approximation (PWIA) calculations \cite{nadia}.
It was pointed out that the spin-dependent part of 
the $NN\to NN^*(\sim 2.2$~GeV/$c^2$) 
amplitude is significant. 
Recent measurements of  the tensor $A_{yy}$ and vector $A_y$ analyzing powers  
in the vicinity of the excitation 
of baryonic resonances with masses up to $\sim 1.8$ GeV/$c^2$ \cite{lad2,lad3} 
demonstrated large spin effects and the sensitivity to the baryonic resonances properties. 
The exclusive measurements   in the
${\rm H}(d,d')X$ reaction in the vicinity of the Roper resonance excitation
performed recently at Saclay \cite{ljuda} have also shown large values  
of the $A_{yy}$ and $A_y$ analyzing powers. 

In this paper we report  new results on the tensor and vector
analyzing powers $A_{yy}$ and $A_y$ in deuteron inelastic scattering
on hydrogen and carbon at the incident deuteron momentum of 9.0~GeV/$c$
and missing masses of undetected system of $\sim$ 2.0--2.6~GeV/$c^2$.
Details
of the experiment are described  in Section~\ref{sec:1}.  The comparison with
 existing
data and theoretical predictions is given in Section~\ref{sec:2}.
Conclusions are drawn in Section~\ref{sec:3}.

\section{\label{sec:1}Experimental procedure}

\begin{figure}
\resizebox{0.5\textwidth}{!}{
\includegraphics{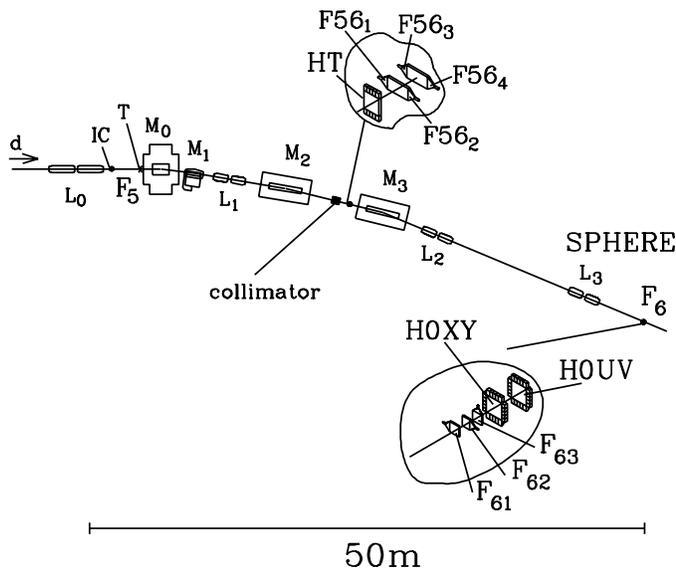}}
\caption{\label{fig:1} Schematic view of the experiment.
$M_i$ and $L_i$ are the magnets and quadrupole
lenses doublets, respectively;
$IC$ is an ionization chamber; $T$ is  a target;
$F_{61},~ F_{62},~F_{63}$  are trigger counters;
$F56_{1-4}$ are scintillation
counters and $HT$ is  a scintillation hodoscope
for TOF measurements;
$H0XY$ and $H0UV$  are beam profile hodoscopes.}
\end{figure}

   The experiment has been performed using the polarized deuteron
beam at the  Dubna Synchrophasotron 
and the SPHERE setup
\cite{lad1,lad2} described elsewhere 
and shown in Fig.\ref{fig:1}.
The polarized deuterons were produced
by the ion source POLARIS \cite{polaris}.
The sign of  beam polarization  changed  cyclically and
spill-by-spill:   $"0"$, $"-"$, $"+"$, where $"0"$ means
the absence of  polarization, $"+"$ and $"-"$ correspond
to the sign of $p_{zz}$ with the quantization axis perpendicular
to the plane containing the mean beam orbit in the accelerator.

The tensor polarization of the beam was
periodically  measured during the experiment using the 
$A(d,p)X$ reaction at 
zero emission angle and a proton momentum of $p_p\sim \frac{2}{3}p_d$ 
\cite{zolin}, using the same setup. It has been
shown that deuteron
breakup reaction in those
kinematical conditions has very large tensor analyzing power
$T_{20}= -0.82\pm 0.04$, independent of the atomic
number of the target
($A >$ 4) and of the momentum of the incident deuterons
between 2.5 and 9.0~GeV/$c$ \cite{t20br}.
The tensor polarization, corrected for 
 dead time effect \cite{dtime}, and averaged over the
whole duration of the experiment   was
$p_{zz}^+=0.798\pm 0.002$(stat.)$\pm 0.040$(syst.) and
$p_{zz}^-=-0.803\pm 0.002$(stat.)$\pm 0.040$(syst.) in $"+"$ and $"-"$
beam spin states, respectively.

The vector polarization of the beam was 
measured from the asymmetry
of quasi-elastic $pp$-scattering on a thin CH$_2$ target
placed   at focus $F4$ about 20~m upstream
of the setup \cite{f4,f4a}. The values of the vector polarization
were obtained using the results of asymmetry
measurements  at a momentum of 4.5~GeV/$c$ per nucleon
and  $8^\circ$  proton scattering angle. The
corresponding value of the effective analyzing power of the polarimeter
$A$(CH$_2$)  amounted $0.123\pm 0.006$ \cite{f4a}.
The vector polarization of the beam in different spin states was
$p_z^+=0.275\pm 0.016$(stat.)$\pm 0.014$(syst.) and
$p_z^-=0.287\pm 0.016$(stat.)$\pm 0.014$(syst.), respectively.

A slowly extracted deuteron beam with a typical intensity 
of $5\times 10^8$ to $10^9$ $\vec{d}$/spill was directed onto
a liquid hydrogen target of 30 cm  
length or onto carbon targets
with different length. 
The beam intensity was monitored by
an ionization chamber placed in front of the target.
The beam positions and profiles at  certain points along the
beam were monitored 
during each spill. The beam size at the target point was
$\sigma_x\sim 0.4$  cm and $\sigma_y\sim  0.9$ cm in the horizontal
and vertical directions, respectively.

Most of the data  were obtained
at secondary deuteron emission angles of 85, 130, and 160 mrad and
deuteron momenta between 4.5 and 7.0 GeV/c on hydrogen and carbon.
The secondary particles emitted  from the target
were transported  to  focus $F_6$ by
means of three bending magnets
and three quadrupole doublets.

    The measurements with empty target were performed 
at secondary particle momenta of 4.5 and 6.0 GeV/$c$, while the measurements 
without  target were carried out  at 4.5~GeV/$c$ only. The
ratio of the yields without target to a 7 cm carbon  
target was less than $\sim 1\%$ for the both 85 mrad and 130 mrad detection angle.
The yield of particles for empty target was
$\sim 3.5\%$ and $\sim 11\%$ of the yield from  30 cm liquid hydrogen target
at 4.5 and 6.0 GeV/$c$, respectively,  independent of emission angle. 

\begin{figure}
\resizebox{0.5\textwidth}{!}{
\includegraphics{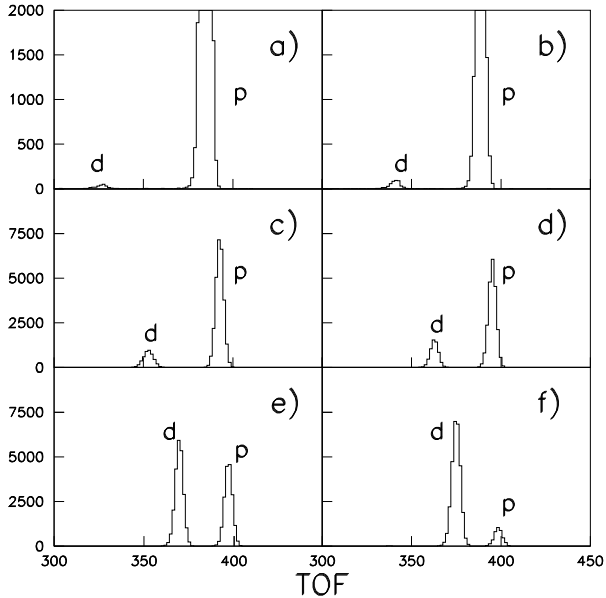}}
\caption{\label{fig:2} TOF spectra (in time-to-digit converter channel numbers) obtained for different setting of the 
magnetic elements. The panels ($a$), ($b$), ($c$), ($d$), ($e$), and ($f$)
correspond to
secondary particles momenta of 4.5, 5.0, 5.5, 6.0, 6.5, and 7.0 GeV/$c$, respectively.}
\end{figure}

The coincidences of signals from 
scintillation counters $F_{61}$, $F_{62}$  and $F_{63}$ were
used as a trigger. Along with the inelastically scattered deuterons, the
apparatus detected the protons originating from deuteron
fragmentation. 
The time-of-flight
(TOF) information with a base line of $\sim$ 34 m between
start counter $F_{61}$ and  stop counters $F56_1$--$F56_2$,
$F56_3$--$F56_4$ and a scintillation hodoscope ${\rm HT}$ was used 
for particle identification
in the off-line  analysis. 
The TOF resolution was better
than $0.2$ ns ($1\sigma$).  The separation of the particles for different momenta of
the detected secondaries is demonstrated in Fig.\ref{fig:2}.  
The panels ($a$), ($b$), ($c$), ($d$), ($e$) and ($f$)
correspond to
secondary particle momenta of
4.5, 5.0, 5.5, 6.0, 6.5, and 7.0 GeV/$c$, respectively.
One can see that the relative contribution of
protons becomes more pronounced with
decreasing momentum.
The residual background was eliminated completely by 
the requirement that particles are detected at least in two prompt 
TOF windows.

The acceptance of the setup was
determined via Monte Carlo simulations, taking into account
the parameters of the incident deuteron beam,
nuclear interactions and multiple scattering  in the target,
in the air, in the windows and detectors, energy losses of primary
and secondary deuterons.
The momentum and polar angle  acceptances were
${\Delta p}/{p}\sim\pm ~2\%$ and $\pm 8$ mrad, respectively.

\begin{figure}
\resizebox{0.5\textwidth}{!}{
\includegraphics{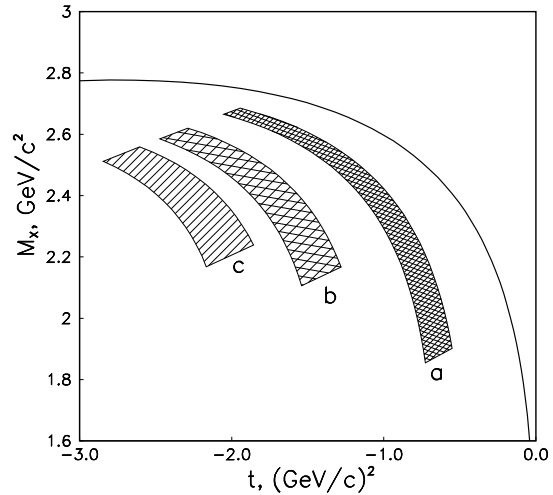}}
\caption{\label{fig:3} The kinematical plot of the missing mass $M_X$ versus 4-momentum
$t$ at the initial deuteron momentum 9.0 GeV/$c$.
The solid line correspond to the conditions (middle of the acceptance)
of the experiment  performed at zero angle \cite{azh9}.
The shadowed areas $a$, $b$, and $c$ demonstrate the regions of 4-momentum $t$ and
missing mass $M_X$ covered  within the acceptance
of the present experiment at the deuteron detection angles 
of $\sim 85$, $\sim 130$, and $\sim 160$ mrad,
respectively.}
\end{figure}

The missing mass values $M_X$  were
calculated under the assumption that the
reaction occurs on a  target with proton mass.
In this case, the 4-momentum transfer $t$ and the missing mass $M_X$ are
related as follows
\begin{eqnarray}
\label{MXt}
M_X^2 = t + m_p^2 + 2m_pQ,
\end{eqnarray}
where $m_p$ is the proton mass and $Q$ is the energy difference
between  incident and scattered deuteron. The
shadowed areas $a$, $b$, and $c$  in the kinematic
plot given in Fig.\ref{fig:3} demonstrate the regions of 4-momentum $t$ and
missing mass $M_X$ covered  by the acceptance setup 
in the present experiment at detection angle of $\sim$85, 130, and 160 mrad, respectively.
The solid line corresponds to an initial deuteron momentum of
9.0 GeV/$c$
and zero emission angle \cite{azh9}. One can see
that the conditions of the present experiment allow one to 
significantly extend 
the $t$ range of measurements for
missing masses $M_X \sim$2.2--2.6~GeV/$c^2$.

The tensor $A_{yy}$ and vector $A_y$ analyzing powers
were calculated from
the numbers of protons $n^+$, $n^-$ and $n^0$, detected for different
states of the beam polarization,
normalized to the corresponding
beam intensities and corrected for the  dead time effect \cite{dtime}.
The calculations were carried out by the expressions
\begin{eqnarray}
\label{ayy}
A_{yy} &=& 2~ \frac{p_z^-~(n^+/n^0-1)
~-~p_z^+~ (n^-/n^0-1)}{p_z^- p_{zz}^+ - p_z^+ p_{zz}^-},\nonumber\\
A_{y} &=& -\frac{2}{3}~
\frac{p_{zz}^-~ (n^+/n^0-1)
~-~p_{zz}^+~ (n^-/n^0-1)}{p_z^- p_{zz}^+ - p_z^+ p_{zz}^-}.
\end{eqnarray}
{These expressions,  taking into account
different values of polarization in different beam spin states,
are simplified significantly when
$p_z^+ = p_z^-$ and $p_{zz}^+ = - p_{zz}^-$.} 
The systematic errors due to the uncertainty of the polarization 
measurements were 
$\sim 5\%$ for both the tensor and vector analyzing powers.

\section{\label{sec:2}Results and discussion}

\begin{table}
\caption{\label{tab:table1} Results on the tensor $A_{yy}$ and vector $A_y$ analyzing powers of the 
${\rm ^1H}(d,d^\prime)X$ reaction
at initial deuteron momentum 9 GeV/$c$.
($\theta$ and $P$  are the polar angle and momentum of the secondary deuteron, respectively; 
$\Delta P$ is momentum acceptance of the setup (FWHM), $t$ is
4-momentum transfer, $M_X$ is missing mass.)}
\begin{ruledtabular}
\begin{tabular}{ccccccc}
$\theta$, & $P$, & $\Delta P$, & $t$, & $M_X$, & $A_{yy}\pm$ &  $A_y\pm$ \\ 
mrad & GeV/$c$ & GeV/$c$ & (GeV/$c$)$^2$ &  GeV/$c^2$  & $\Delta A_{yy}$ & $\Delta A_y$\\
\hline 
0.085 & 4.541 & 0.095 & -1.875 & 2.651 & -0.110$\pm$ &  0.037$\pm$ \\ 
      &       &       &        &       &  0.294~~    &  0.283~~ \\ 
0.085 & 5.041 & 0.105 & -1.466 & 2.562 & -0.013$\pm$ & -0.183$\pm$\\ 
      &       &       &        &       &  0.074~~    &  0.196~~ \\ 
0.085 & 6.047 & 0.125 & -0.940 & 2.302 &  0.066$\pm$ &  0.189$\pm$ \\ 
      &       &       &        &       &  0.065~~    &  0.062~~ \\ 
0.085 & 6.554 & 0.137 & -0.782 & 2.132 &  0.053$\pm$ &  0.174$\pm$\\
      &       &       &        &       &  0.041~~    &  0.039~~ \\ 
0.130 & 4.555 & 0.096 & -2.246 & 2.575 &  0.061$\pm$ & -0.277$\pm$ \\ 
      &       &       &        &       &  0.247~~    &  0.235~~ \\ 
0.130 & 5.051 & 0.109 & -1.883 & 2.475 &  0.042$\pm$ & -0.092$\pm$ \\ 
      &       &       &        &       &  0.165~~    &  0.156~~ \\ 
0.130 & 6.061 & 0.130 & -1.446 & 2.184 & -0.024$\pm$ &  0.101$\pm$\\
      &       &       &        &       &  0.047~~    &  0.045~~ \\ 
\end{tabular}
\end{ruledtabular}
\end{table}

The data on $A_{yy}$ and $A_y$ with the statistical errors for 
the ${\rm ^1H}(d,d^\prime)X$ and  ${\rm ^{12}C}(d,d^\prime)X$ are presented
in the Tables {\ref{tab:table1} and {\ref{tab:table2}, respectively.

\begin{figure}
\resizebox{0.5\textwidth}{!}{
\includegraphics{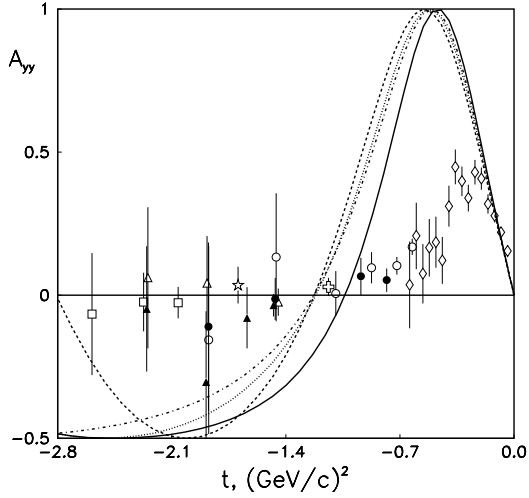}}
\caption{\label{fig:4} Tensor analyzing power $A_{yy}$ in the $(d,d^\prime)X$ reaction obtained at  9
GeV/$c$. The symbols are explained in the text. The solid, dashed, dotted
and dash-dotted lines are predictions in the framework of
PWIA \cite{nadia} using
DWFs for Paris \cite{paris} and Bonn A, B and C \cite{bonn}
potentials, respectively.}
\end{figure}

The results of the present experiment on the tensor analyzing power $A_{yy}$ 
in the $(d,d^\prime)X$ reaction are shown in Fig.\ref{fig:4} as a function of the 
transferred 4-momentum  $t$.  
The open circles, triangles and squares are the data obtained on a carbon target
at detection angles of $\sim 85$,
$\sim 130$, and $\sim 160$ mrad, respectively. The cross and star correspond to the $A_{yy}$
data for a carbon target taken at $\sim 115$ and $\sim 145$ mrad.
The filled circles and triangles denote
 the data obtained on a hydrogen target at $\sim 85$ and
$\sim 130$ mrad, respectively.     The data of the 
present experiment are compared with
the data taken at zero angle on hydrogen~\cite{azh9} given by the diamonds.
One can see that the values of $A_{yy}$ obtained at different emission angles are in good 
agreement demonstrating an approximate $t$-scaling observed in
 earlier experiments 
\cite{45_55}-\cite{lad3}.
The present data along with the data from previous experiments
allow one to trace the general behaviour
of $A_{yy}$   over a wide region of $|t|$.
At small $|t|$ ($\le 0.3$~(GeV/$c$)$^2$), $A_{yy}$ 
rises linearly up to a 
value of $\sim 0.4$,
then it smoothly decreases towards zero at $|t|\sim$ 1.0--1.4~ (GeV/$c$)$^2$.

\begin{table}
\caption{\label{tab:table2} Results on the tensor $A_{yy}$ and vector $A_y$ analyzing powers of the 
${\rm ^{12}C}(d,d^\prime)X$  reaction
at initial deuteron momentum 9 GeV/$c$.
($\theta$ and $P$ are the polar angle and momentum of the secondary deuteron, respectively; 
$\Delta P$ is  momentum acceptance of the setup (FWHM), $t$ is 
4-momentum transfer, $M_X$ is missing mass.)}
\begin{ruledtabular}
\begin{tabular}{ccccccc}
$\theta$, & $P$, & $\Delta P$, & $t$, & $M_X$, & $A_{yy}\pm$ &  $A_y\pm$ \\ 
mrad & GeV/$c$ & GeV/$c$ & (GeV/$c$)$^2$ &  GeV/$c^2$  & $\Delta A_{yy}$ & $\Delta A_y$\\
\hline 
0.085 & 4.544 & 0.089 & -1.873 & 2.650 & -0.156$\pm$ &  0.470$\pm$ \\
      &       &       &        &       &  0.329~~    &  0.314~~ \\
0.085 & 5.050 & 0.098 & -1.460 & 2.559 &  0.133$\pm$ &  0.264$\pm$ \\
      &       &       &        &       &  0.224~~    &  0.213~~ \\
0.085 & 5.606 & 0.105 & -1.094 & 2.429 &  0.006$\pm$ &  0.161$\pm$ \\
      &       &       &        &       &  0.079~~    &  0.075~~ \\
0.085 & 6.109 & 0.116 & -0.875 & 2.285 &  0.096$\pm$ & -0.070$\pm$ \\
      &       &       &        &       &  0.054~~    &  0.051~~ \\
0.085 & 6.616 & 0.126 & -0.719 & 2.113 &  0.103$\pm$ &  0.134$\pm$ \\
      &       &       &        &       &  0.031~~    &  0.030~~ \\
0.085 & 7.117 & 0.131 & -0.625 & 1.911 &  0.169$\pm$ &  0.120$\pm$ \\
      &       &       &        &       &  0.029~~    &  0.027~~ \\
0.115 & 6.587 & 0.131 & -1.138 & 2.024 &  0.029$\pm$ &  0.147$\pm$\\ 
      &       &       &        &       &  0.032~~    &  0.030~~ \\   
0.130 & 4.563 & 0.087 & -2.254 & 2.570 & -0.048$\pm$ &  0.100$\pm$\\
      &       &       &        &       &  0.219~~    &  0.208~~ \\
0.130 & 5.070 & 0.096 & -1.889 & 2.467 & -0.303$\pm$ &  0.298$\pm$ \\
      &       &       &        &       &  0.247~~    &  0.235~~ \\ 
0.130 & 5.590 & 0.107 & -1.638 & 2.321 & -0.079$\pm$ &  0.257$\pm$ \\
      &       &       &        &       &  0.107~~    &  0.102~~ \\
0.130 & 6.094 & 0.116 & -1.476 & 2.156 & -0.035$\pm$ &  0.023$\pm$ \\
      &       &       &        &       &  0.039~~    &  0.037~~ \\
0.145 & 6.097 & 0.118 & -1.694 & 2.104 &  0.035$\pm$ &  0.087$\pm$ \\
      &       &       &        &       &  0.064~~    &  0.061~~ \\
0.160 & 4.590 & 0.091 & -2.589 & 2.490 & -0.066$\pm$ &  0.275$\pm$ \\
      &       &       &        &       &  0.214~~    &  0.204~~ \\
0.160 & 5.095 & 0.097 & -2.272 & 2.373 & -0.024$\pm$ &  0.155$\pm$ \\
      &       &       &        &       &  0.102~~    &  0.097~~ \\
0.160 & 5.600 & 0.109 & -2.060 & 2.225 & -0.026$\pm$ &  0.044$\pm$ \\
      &       &       &        &       &  0.055~~    &  0.053~~ \\
\end{tabular}
\end{ruledtabular}
\end{table}

The data obtained on hydrogen and carbon targets are in good
agreement.   
The observed independence of the tensor analyzing power on the
atomic number of the target indicates that rescattering in
the target and medium effects  are small. Hence, nuclear targets
are  appropriate to obtain information  on baryonic
excitations in inelastic  deuteron scattering.

\begin{figure}
\resizebox{0.5\textwidth}{!}{
\includegraphics{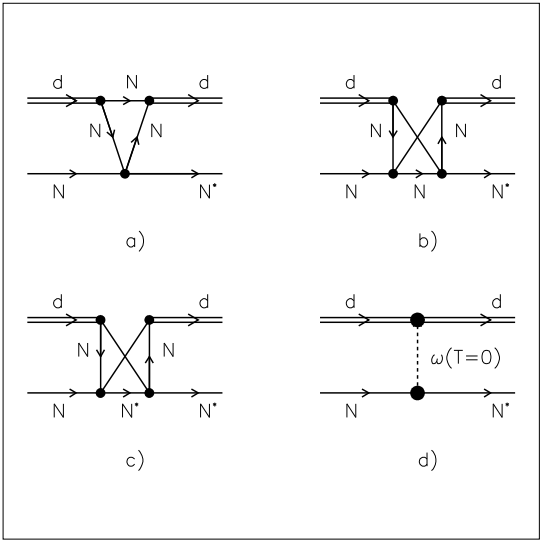}}
\caption{\label{fig:4a} Diagrams representing mechanisms of baryonic excitation
in  deuteron--nucleon collisions:
($a$) plane wave impulse approximation; ($b$) and 
($c$) double-collision mechanism; ($d$)
$\omega$-meson exchange mechanism.}
\end{figure}

The results on the tensor $A_{yy}$ and vector $A_y$ obtained at 9 GeV/$c$ and 85 mrad\cite{lad2}
 and at 5 GeV/$c$ and 178 mrad \cite{lad3} have been satisfactorily explained in the framework of the
PWIA approach \cite{nadia} (see diagram in Fig.~\ref{fig:4a}$a$).
In this model the tensor and vector analyzing powers are expressed in terms of
three amplitudes ($T_{00}$, $T_{11}$ and $T_{10}$) defined by the deuteron 
structure and  the ratio $r$ of the spin-dependent to spin-independent 
parts of the elementary process $NN\to NN^*$ in the vicinity of $\sim 2.2$ GeV/$c^2$:
\begin{eqnarray}
\label{ayyr}
A_{yy}(q) &=& \frac{T_{00}^2 - T_{11}^2 +  4 r^2 T_{10}^2}
{T_{00}^2 + 2 T_{11}^2 +  4 r^2 T_{10}^2},\\
\label{ayr}
A_y(q) &=& 2\sqrt{2} r \frac{(T_{11}+T_{00})T_{10}}
{T_{00}^2 + 2 T_{11}^2 +  4 r^2 T_{10}^2}.
\end{eqnarray}

Amplitudes 
$T_{00}$ and  $T_{11}$
depend on the deuteron wave function:
\begin{eqnarray}
\label{t00}
T_{00} &=& S_0(q/2)+{\sqrt{2}}S_2(q/2),\nonumber\\
T_{11} &=& S_0(q/2)-\frac{1}{\sqrt{2}}S_2(q/2),
\end{eqnarray}
where $S_0$ and $S_2$ are the spherical and quadrupole form factors
of the deuteron expressed in the following manner:
\begin{eqnarray}
\label{quad}
S_0(q/2) & = & \int^{\infty}_0 (u^2(r) + w^2(r)) j_0(rq/2) dr, \nonumber\\
S_2(q/2) & = & \int^{\infty}_{0} {2} w(r) \left ( u(r) - \frac{1}{2\sqrt{2}}w(r) \right )
j_2(rq/2) dr.
\end{eqnarray}
Here, $u(r)$ and $w(r)$  are $S$-- and  $D$-- 
waves of the deuteron in coordinate space, respectively,
$j_0(qr/2)$ and $j_2(qr/2)$ 
are the Bessel functions of  zero and second order, respectively, and $q^2 = -t$.

The amplitude $T_{10}$ 
is also function of $S$-- and $D$-- waves of the deuteron
\begin{eqnarray} 
\label{t10}
T_{10} = & & \frac{{\it i}}{\sqrt{2}}
\int^{\infty}_0 \left (u^2(r) - \frac{w^2(r)}{2}\right ) j_0(rq/2)
dr+\nonumber\\
+& &\frac{{\it i}}{2}\int^{\infty}_{0} w(r) \left ( u(r) +
\frac{w(r)}{\sqrt{2}} \right )
j_2(rq/2) dr .
\end{eqnarray}
One can see that in the framework of the PWIA model \cite{nadia},
the vector analyzing power $A_y$ is proportional to  the ratio $r$,
while the tensor analyzing power $A_{yy}$ is defined mostly by the deuteron structure.

\begin{figure}
\resizebox{0.5\textwidth}{!}{
\includegraphics{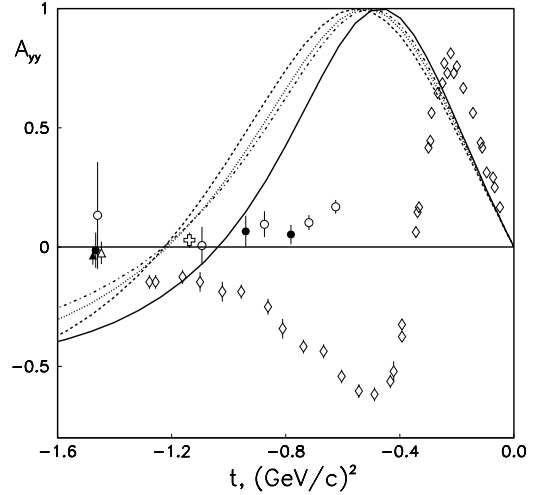}}
\caption{\label{fig:6} Comparison of the tensor analyzing power $A_{yy}$ in the $(d,d^\prime)X$ reaction 
from the present experiment with the $A_{yy}$ data in deuteron-proton elastic scattering 
at 3.43 GeV/c \cite{blez} (diamonds).
The symbols for the $(d,d^\prime)X$ reaction data and curves are 
the same as in Fig.\ref{fig:4}.}
\end{figure}

The curves shown in Fig.\ref{fig:4} 
are the results of PWIA calculations \cite{nadia} with different
deuteron wave functions. The solid line 
is calculated with the deuteron wave function (DWF) for the Paris 
potential \cite{paris},
while  the dashed, dotted and dash-dotted lines
correspond to the DWFs for Bonn A, B and C potentials \cite{bonn},
respectively.

The deviation of
the data obtained in the previous experiments \cite{45_55,lad2}
at $|t|\sim$~ 0.3--0.8~(GeV/$c$)$^2$ 
from the predictions of PWIA, as well as from 
$dp$- \cite{blez} and $ed$- \cite{ed} elastic scattering,
indicates the sensitivity of $A_{yy}$ to the
baryonic resonances excitation.
In the framework of the multiple-scattering model, such a deviation
may be due to a significant contribution of double-collision
interactions
corresponding to the diagrams shown in Figs.\ref{fig:4a}$b$ and \ref{fig:4a}$c$.
The diagram of Fig.\ref{fig:4a}$b$ describes the situation, where the
resonance is formed in the second $NN$ collision whereas
the resonance in the case of Fig.\ref{fig:4a}$c$,
formed in the first $NN$ interaction, is
then elastically scattered  on  the second nucleon
of the deuteron. The calculations have shown that the contribution
of the double-scattering terms
is significant for $|t|$ greater than $\sim$ 0.4 (GeV/$c$)$^2$
\cite{azhgirey2}. Therefore, in the framework of the multiple-scattering
model, the behaviour of the tensor analyzing power $A_{yy}$ is defined
by the spin structure of the deuteron and the elementary amplitudes
of the $NN\to NN^*$ and $NN^*\to NN^*$ processes.

     The sensitivity of the tensor analyzing power in deuteron
 inelastic scattering off protons to the
excitation of baryonic resonances is 
pointed out   in the framework of the $t$-channel
$\omega$-meson  exchange model \cite{egle1}. The diagram corresponding
to this model is shown in Fig.\ref{fig:4a}$d$.  The cross section and
polarization observables can be calculated from the known
electromagnetic properties of the deuteron and baryonic resonances
$N^*$ through the vector dominance model. The details of the
model are given in \cite{egle1,egle2}.
This model satisfactorily explained  the $A_{yy}$ data in the $(d,d^\prime)X$ reaction in the vicinity of low masses baryonic resonances 
\cite{45_55,azh9,lad2}.
In principle, this model can be applied for the explanation of the data of 
present experiment, however, the knowledge of the electromagnetic properties
of the baryonic resonances with the masses 2.2--2.6 ~ GeV/$c^2$
is required.

Fig. \ref{fig:6} demonstrates 
the different behavior of the tensor analyzing powers in the $(d,d^\prime)X$
process obtained in the present experiment and in 
$dp$- \cite{blez} elastic scattering, indicating different contribution of
the double scattering at large $|t|$. 
One can see also, that the  present $A_{yy}$ data are in reasonable agreement with
the PWIA prediction in the range of transferred 4-momentum $|t|$ of 0.9--1.2 
(GeV/$c$)$^2$.

\begin{figure}
\resizebox{0.5\textwidth}{!}{
\includegraphics{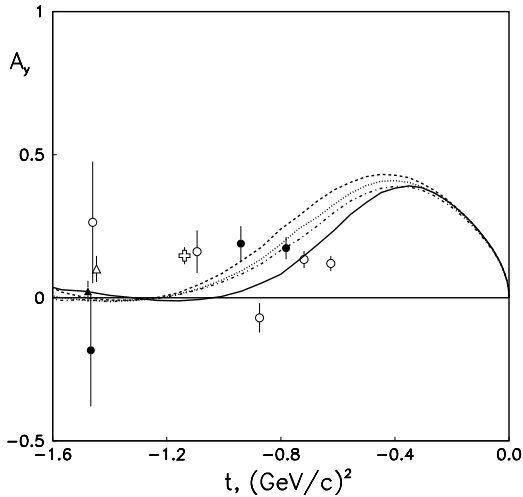}}
\caption{\label{fig:7} Vector analyzing power $A_{y}$ in the $(d,d^\prime)X$ reaction 
from the present experiment compared with the 
PWIA predictions \cite{nadia} using different DWFs.
}
\end{figure}

The data on the vector analyzing power $A_y$ obtained in 
the present experiment are plotted in Fig.\ref{fig:7} versus $t$.
The symbols corresponding to the data obtained in different
kinematic conditions are the same as in Fig.\ref{fig:4}. 
It should be noted that
the data obtained on hydrogen and carbon are in good agreement.
The curves are obtained using the expression (\ref{ayr}) with 
the ratio $r$ of the
spin-dependent to spin-independent parts of the $NN\to NN^*$ process
taken in the form 
\begin{eqnarray}
\label{r04}
r=a\cdot q
\end{eqnarray}
with the value of $a=0.4$ \cite{nadia}.
The solid curve in Fig.\ref{fig:7}
is obtained with the DWF for the Paris potential \cite{paris},
while  the dashed, dotted and dash-dotted lines
correspond to the DWFs for Bonn A, B and C potentials \cite{bonn},
respectively. 
The PWIA calculations give approximately the same results 
at the value of $a\sim$~0.3--0.5.  
It should be noted that $a$  might 
be different for different $M_X$, however,
we took the fixed value  for simplicity
due to the lack of data. 
The observed relatively large value of $A_y$ at $|t|\sim$
1~ (GeV/$c$)$^2$ can be interpreted as being due to  a significant
role of the spin-dependent part of the elementary amplitude of the 
$NN\to NN^*$ reaction.

\section{\label{sec:3}Conclusion}

     We have presented data on the tensor and vector analyzing
powers $A_{yy}$ and $A_y$
in inelastic scattering $(d,d^\prime)X$ of 9.0 GeV/$c$ deuterons
on hydrogen and carbon 
in the vicinity of
the excitation of baryonic masses of
 $\sim$2.2--2.6 GeV/$c^2$. This corresponds to the range of 4-momentum
$|t|$ between 0.7 and 2.5 (GeV/$c$)$^2$.

The data obtained on hydrogen and carbon targets are in good
agreement.   
The observed independence of the analyzing powers on the
atomic number of the target indicates that rescattering in
the target and medium effects  are small. Hence, nuclear targets
are also appropriate to obtain information  on the baryonic
excitation in 
inelastic deuteron scattering \cite{45_55}-\cite{lad3}.

The observed difference of the $A_{yy}$ data 
at $|t|\le$ 0.8~(GeV/$c$)$^2$
with the PWIA calculations \cite{nadia} using
standard deuteron wave functions as well as with the data obtained in $dp$- \cite{blez} and  $ed$-elastic scattering \cite{ed} indicates the sensitivity of $A_{yy}$ to
baryonic resonance excitation via 
double-collision
interactions \cite{azhgirey2}. On the other hand, both $A_{yy}$ and $A_y$ data 
are in reasonable 
agreement with PWIA calculations \cite{nadia} at $|t|\sim$  1 ~(GeV/$c$)$^2$.

The large values of $A_y$ show a significant
role of the spin-dependent part of the elementary amplitude of the 
$NN\to NN^*$ reaction in the vicinity of the baryonic masses excitation  $M_X\sim$2.2~GeV/$c^2$.

\begin{acknowledgments}

Authors express their gratitude to  Prof.~A.I.Malakhov, Director of
LHE, and Prof.~V.N.Penev for their permanent help.
Authors are grateful to the LHE accelerator staff and POLARIS team
for providing good conditions for the
experiment.
They thank to
L.V.~Budkin, V.P.~Ershov,
V.V.~Fimushkin,
A.S.~Nikiforov,
Yu.K.~Pilipenko,
V.G.~Perevozchikov,  E.V.~Ryzhov, A.I.~Shirokov, and O.A.~Titov  for their
assistance during the experiment.
Authors thank also A.N.~Prokofiev and A.A.~Zhdanov for the lending
of the scintillation counters for the polarimeter.
They are indebted to  Prof.~E.Tomasi-Gustafsson  for helpful
discussions.
This work was supported in part by the Russian Foundation
for Basic Research under grants
No.~03-02-16224 and No.~04-02-17107.
\end{acknowledgments}



\begin{thebibliography}{99}
\expandafter\ifx\csname natexlab\endcsname\relax\def\natexlab#1{#1}\fi
\expandafter\ifx\csname bibnamefont\endcsname\relax
  \def\bibnamefont#1{#1}\fi
\expandafter\ifx\csname bibfnamefont\endcsname\relax
  \def\bibfnamefont#1{#1}\fi
\expandafter\ifx\csname citenamefont\endcsname\relax
  \def\citenamefont#1{#1}\fi
\expandafter\ifx\csname url\endcsname\relax
  \def\url#1{\texttt{#1}}\fi
\expandafter\ifx\csname urlprefix\endcsname\relax\def\urlprefix{URL }\fi
\providecommand{\bibinfo}[2]{#2}
\providecommand{\eprint}[2][]{\url{#2}}

\bibitem[{\citenamefont{L.S.Azhgirey~et~al.}(1995)}]{45_55}
\bibinfo{author}{\bibfnamefont{L.S.~}\bibnamefont{Azhgirey~et~al.}},
  \bibinfo{journal}{Phys.~ Lett.~B} \textbf{\bibinfo{volume}{361}},
  \bibinfo{pages}{21} (\bibinfo{year}{1995}).

\bibitem[{\citenamefont{L.S.Azhgirey~et~al.}(1998)}]{azh9}
\bibinfo{author}{\bibfnamefont{L.S.~}\bibnamefont{Azhgirey~et~al.}},
  \bibinfo{journal}{JINR~ Rapid.~ Comm.~N$^o$} \textbf{\bibinfo{volume}{2[88]-98}},
  \bibinfo{pages}{17} (\bibinfo{year}{1998}).

\bibitem[{\citenamefont{L.S.Azhgirey~et~al.}(1999)}]{lad1}
\bibinfo{author}{\bibfnamefont{L.S.~}\bibnamefont{Azhgirey~et~al.}},
  \bibinfo{journal}{Yad.~Fiz.} \textbf{\bibinfo{volume}{62}},
  \bibinfo{pages}{1796} (\bibinfo{year}{1999})
  [\bibinfo{journal}{Phys.~Atom.~Nucl.}
  \textbf{\bibinfo{volume}{62}}, \bibinfo{pages}{1673}
  (\bibinfo{year}{1999})].

\bibitem[{\citenamefont{V.P.Ladygin et~al.}(2000)}]{lad2}
\bibinfo{author}{\bibfnamefont{V.P.~}\bibnamefont{Ladygin et~al.}},
  \bibinfo{journal}{Eur.~Phys.~J.~A} \textbf{\bibinfo{volume}{8}},
  \bibinfo{pages}{409} (\bibinfo{year}{2000});
\bibinfo{author}{\bibfnamefont{L.S.~}\bibnamefont{Azhgirey~et~al.}},
  \bibinfo{journal}{Yad.~Fiz.} \textbf{\bibinfo{volume}{64}},
  \bibinfo{pages}{2046} (\bibinfo{year}{2001})  
  [\bibinfo{journal}{Phys.~Atom.~Nucl.}
  \textbf{\bibinfo{volume}{64}}, \bibinfo{pages}{1961}
  (\bibinfo{year}{2001})].

\bibitem[{\citenamefont{L.S.~Azhgirey et~al.}(2005)}]{lad3}
\bibinfo{author}{\bibfnamefont{L.S.~}\bibnamefont{Azhgirey~et~al.}},
  \bibinfo{journal}{Yad.~Fiz.} \textbf{\bibinfo{volume}{68}},
  \bibinfo{pages}{1029} (\bibinfo{year}{2005}) 
  [\bibinfo{journal}{Phys.~Atom.~Nucl.}
  \textbf{\bibinfo{volume}{68}}, \bibinfo{pages}{991}
  (\bibinfo{year}{2005})].

\bibitem[{\citenamefont{M.~Morlet et~al.}(1994)}]{morlet}
\bibinfo{author}{\bibfnamefont{}~\bibnamefont{Experiment LNS-E250 (unpublished)}}

\bibitem[{\citenamefont{L.V.Malinina et~al.}(2001)}]{ljuda}
\bibinfo{author}{\bibfnamefont{L.V.~}\bibnamefont{Malinina~et~al.}},
  \bibinfo{journal}{Phys.~Rev.~C} \textbf{\bibinfo{volume}{64}},
  \bibinfo{pages}{064001} (\bibinfo{year}{2001}). 

\bibitem[{\citenamefont{T.~Suzuki}(1994)}]{suzuki}
\bibinfo{author}{\bibfnamefont{T.~}\bibnamefont{Suzuki}},
  \bibinfo{journal}{Nucl.~Phys.~A} \textbf{\bibinfo{volume}{577}},
  \bibinfo{pages}{167} (\bibinfo{year}{1994}). 

\bibitem[{\citenamefont{Y.~Satou et al.}(2001)}]{satou}
\bibinfo{author}{\bibfnamefont{Y.~}\bibnamefont{Satou et al.}},
  \bibinfo{journal}{Phys.~Lett.~B} \textbf{\bibinfo{volume}{521}},
  \bibinfo{pages}{153} (\bibinfo{year}{2001}). 

\bibitem[{\citenamefont{E.~Tomasi-Gustafsson}(1996)}]{egle1}
\bibinfo{author}{\bibfnamefont{M.P.~}\bibnamefont{Rekalo}} \bibnamefont{and}
\bibinfo{author}{\bibfnamefont{E.~}\bibnamefont{Tomasi-Gustafsson}}, 
 \bibinfo{journal}{Phys.~Rev.~C} \textbf{\bibinfo{volume}{54}},
  \bibinfo{pages}{3125} (\bibinfo{year}{1996}). 

\bibitem[{\citenamefont{E.~Tomasi-Gustafsson}(1999)}]{egle2}
\bibinfo{author}{\bibfnamefont{E.~}\bibnamefont{Tomasi-Gustafsson}},
\bibinfo{author}{\bibfnamefont{M.P.~}\bibnamefont{Rekalo}},
\bibinfo{author}{\bibfnamefont{R.~}\bibnamefont{Bijker}}\bibnamefont{et al.},
 \bibinfo{journal}{Phys.~Rev.~C} \textbf{\bibinfo{volume}{59}},
  \bibinfo{pages}{1526} (\bibinfo{year}{1999}).

\bibitem[{\citenamefont{N.B.~Ladygina}(2002)}]{nadia}
\bibinfo{author}{\bibfnamefont{V.P.}\bibnamefont{Ladygin}} \bibnamefont{and}
\bibinfo{author}{\bibfnamefont{N.B.}\bibnamefont{Ladygina}}, 
   \bibinfo{journal}{Yad.~Fiz.} \textbf{\bibinfo{volume}{65}},
  \bibinfo{pages}{188} (\bibinfo{year}{2002}) 
  [\bibinfo{journal}{Phys.~Atom.~Nucl.}
  \textbf{\bibinfo{volume}{65}}, \bibinfo{pages}{182}
  (\bibinfo{year}{2002})].



\bibitem[{\citenamefont{N.G.~Anishchenko,}(1982)}]{polaris}
\bibinfo{author}{\bibfnamefont{N.G.~}\bibnamefont{Anishchenko et al.}},
   \bibinfo{journal}{in Proceedings of the 5-th International Symposium 
on High Energy Spin Physics (Brookhaven, 1982), AIP Conf.Proc.} 
   \textbf{\bibinfo{volume}{95}},
  \bibinfo{pages}{445} (\bibinfo{year}{1983}).

\bibitem[{\citenamefont{L.S.~Zolin~et~al.}(1998)}]{zolin}
\bibinfo{author}{\bibfnamefont{L.S.~}\bibnamefont{Zolin~et~al.}},
  \bibinfo{journal}{JINR~Rapid.~Comm.~N$^o$} \textbf{\bibinfo{volume}{2[88]-98}},
  \bibinfo{pages}{27} (\bibinfo{year}{1998}).

\bibitem[{\citenamefont{C.F.~Perdrisat et al.}(2001)}]{t20br}
\bibinfo{author}{\bibfnamefont{C.F.~}\bibnamefont{Perdrisat et al.}},
  \bibinfo{journal}{Phys.~Rev.~Lett.} \textbf{\bibinfo{volume}{59}},
  \bibinfo{pages}{2840} (\bibinfo{year}{1987}); 
\bibinfo{author}{\bibfnamefont{V.~}\bibnamefont{Punjabi et al.}},
  \bibinfo{journal}{Phys.~Rev.~C} \textbf{\bibinfo{volume}{39}},
  \bibinfo{pages}{608} (\bibinfo{year}{1989});
\bibinfo{author}{\bibfnamefont{V.G.~}\bibnamefont{Ableev et al.}},
  \bibinfo{journal}{Pis'ma~Zh.~Eksp.~Teor.~Fiz.} \textbf{\bibinfo{volume}{47}},
  \bibinfo{pages}{558} (\bibinfo{year}{1988});
\bibinfo{author}{\bibfnamefont{V.G.~}\bibnamefont{Ableev et al.}},
  \bibinfo{journal}{JINR~ Rapid~Comm.~N$^o$} \textbf{\bibinfo{volume}{4[43]-90}},
  \bibinfo{pages}{5} (\bibinfo{year}{1990});
\bibinfo{author}{\bibfnamefont{T.~}\bibnamefont{Aono et al.}},
  \bibinfo{journal}{Phys.~Rev.~Lett.} \textbf{\bibinfo{volume}{74}},
  \bibinfo{pages}{4997} (\bibinfo{year}{1995}).

\bibitem[{\citenamefont{V.P.~Ladygin}(1999)}]{dtime}
\bibinfo{author}{\bibfnamefont{V.P.~}\bibnamefont{Ladygin}}, 
   \bibinfo{journal}{Nucl.~Instrum.~Methods~A} 
   \textbf{\bibinfo{volume}{437}},
  \bibinfo{pages}{98} (\bibinfo{year}{1999}).

\bibitem[{\citenamefont{L.S.~Azhgirey et~al.}(1997)}]{f4}
\bibinfo{author}{\bibfnamefont{L.S.~}\bibnamefont{Azhgirey~et~al.}},
  \bibinfo{journal}{Prib.~Tekhnik.~Eksp.} \textbf{\bibinfo{volume}{1}},
  \bibinfo{pages}{57} (\bibinfo{year}{1997}) 
  [\bibinfo{journal}{Instr.~Exp.~Tech.}
  \textbf{\bibinfo{volume}{40}}, \bibinfo{pages}{43}
  (\bibinfo{year}{1997})].

\bibitem[{\citenamefont{L.S.~Azhgirey et~al.}(2003)}]{f4a}
\bibinfo{author}{\bibfnamefont{L.S.~}\bibnamefont{Azhgirey~et~al.}},
   \bibinfo{journal}{Nucl.~Instrum.~Methods~A} 
   \textbf{\bibinfo{volume}{497}},
  \bibinfo{pages}{340} (\bibinfo{year}{2003}).

\bibitem[{\citenamefont{M.~Lacombe et~al.}(1981)}]{paris}
\bibinfo{author}{\bibfnamefont{M.~}\bibnamefont{Lacombe~et al.}},
   \bibinfo{journal}{Phys.~Lett.~B} 
   \textbf{\bibinfo{volume}{101}},
  \bibinfo{pages}{139} (\bibinfo{year}{1981}).

\bibitem[{\citenamefont{R.~Machleidt et al.}(1987)}]{bonn}
\bibinfo{author}{\bibfnamefont{R.~}\bibnamefont{Machleidt et al.}},
   \bibinfo{journal}{Phys.~Rep.} 
   \textbf{\bibinfo{volume}{149}},
  \bibinfo{pages}{1} (\bibinfo{year}{1987}).

\bibitem[{\citenamefont{M.~Bleszynski, et al.,}(1979)}]{blez}
\bibinfo{author}{\bibfnamefont{M.~}\bibnamefont{Bleszynski et al.}},
   \bibinfo{journal}{Phys.~Lett.~B} 
   \textbf{\bibinfo{volume}{87}},
  \bibinfo{pages}{178} (\bibinfo{year}{1979});
\bibinfo{author}{\bibfnamefont{M.~}\bibnamefont{Haji-Saied et al.}},
   \bibinfo{journal}{Phys.~Rev.~C} 
   \textbf{\bibinfo{volume}{36}},
  \bibinfo{pages}{2010} (\bibinfo{year}{1987}).

\bibitem[{\citenamefont{M.~Garcon et al.,}(1979)}]{ed}
\bibinfo{author}{\bibfnamefont{M.~}\bibnamefont{Garcon et al.}},
   \bibinfo{journal}{Phys.~Rev.~C} 
   \textbf{\bibinfo{volume}{49}},
  \bibinfo{pages}{2516} (\bibinfo{year}{1994});
\bibinfo{author}{\bibfnamefont{D.~}\bibnamefont{Abbott et al.}},
   \bibinfo{journal}{Phys.~Rev.~Lett.} 
   \textbf{\bibinfo{volume}{84}},
  \bibinfo{pages}{5053} (\bibinfo{year}{2000});
\bibinfo{author}{\bibfnamefont{D.M.~}\bibnamefont{Nikolenko et al.}},
   \bibinfo{journal}{Phys.~Rev.~Lett.} 
   \textbf{\bibinfo{volume}{90}},
  \bibinfo{pages}{072501} (\bibinfo{year}{2003}).

\bibitem[{\citenamefont{L.S.~Azhgirey et~al.}(1988)}]{azhgirey2}
\bibinfo{author}{\bibfnamefont{L.S.~}\bibnamefont{Azhgire\u\i~et~al.}},
  \bibinfo{journal}{Yad.~Fiz.} \textbf{\bibinfo{volume}{48}},
  \bibinfo{pages}{1758} (\bibinfo{year}{1988}) 
  [\bibinfo{journal}{Sov.~J.~Nucl.~Phys.}
  \textbf{\bibinfo{volume}{48}}, \bibinfo{pages}{1058}
  (\bibinfo{year}{1988})].
\end{thebibliography}

\end{document}